\documentclass[aps,twocolumn,floats,floatfix,superscriptaddress,showpacs]{revtex4}

\usepackage{graphicx,amsmath,amssymb,multirow,rotate,color}
\usepackage{ae,aecompl}

\usepackage{color}
\definecolor{red}{rgb}{1,0,0}
\definecolor{green}{rgb}{0,1,0}
\definecolor{blue}{rgb}{0,0,1}
\newcommand{\xxr}[1]{\textcolor{red}{#1}}

\begin{document}

\title{Self-similarities in the  frequency-amplitude space of 
       a loss-modulated CO$_2$ laser}

\author{Cristian \surname{Bonatto}}
\affiliation{Instituto de F\'\i sica, Universidade Federal do 
Rio Grande do Sul, 91501-970 Porto Alegre, Brazil}
\author{Jean Claude \surname{Garreau}}
\affiliation{Laboratoire de Physique des Lasers, Atomes et
  Mol\'ecules, Centre d'\'Etudes et de Recherches Laser et Applications,
  Universit\'e des Sciences et Technologies de Lille,
  F-59655 Villeneuve d'Ascq CEDEX, France}
\author{Jason A.C.~\surname{Gallas}}
\affiliation{Instituto de F\'\i sica, 
             Universidade Federal do Rio Grande do Sul, 
             91501-970 Porto Alegre, Brazil}
\affiliation{Laboratoire de Physique des Lasers, Atomes et
  Mol\'ecules, Centre d'\'Etudes et de Recherches Laser et Applications,
  Universit\'e des Sciences et Technologies de Lille,
  F-59655 Villeneuve d'Ascq CEDEX, France}

\date{\today}

\begin{abstract}
We show the standard two-level continuous-time model of loss-modulated CO$_2$ 
lasers to display the same regular network of self-similar 
stability islands known so far to be typically present only in 
discrete-time models based on mappings. 
For class B laser models our results suggest that, 
more than just convenient surrogates, 
discrete mappings in fact could be isomorphic to continuous flows.
\end{abstract}


\maketitle

\begin{figure*}[bht]
\includegraphics[width=5.7cm]{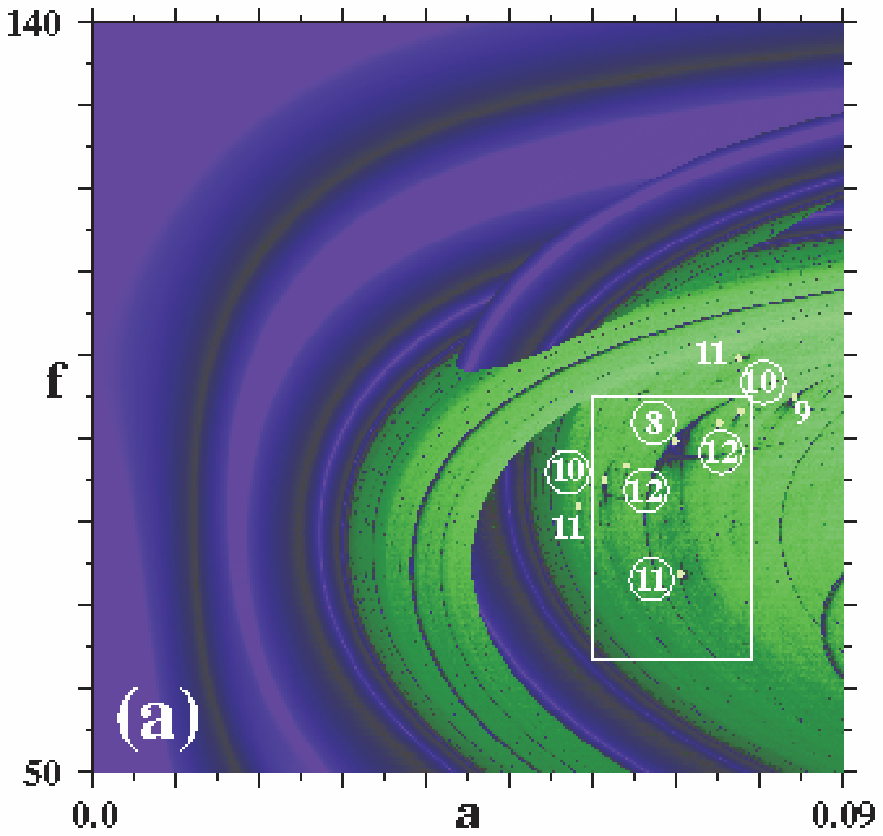}%
\includegraphics[width=5.7cm]{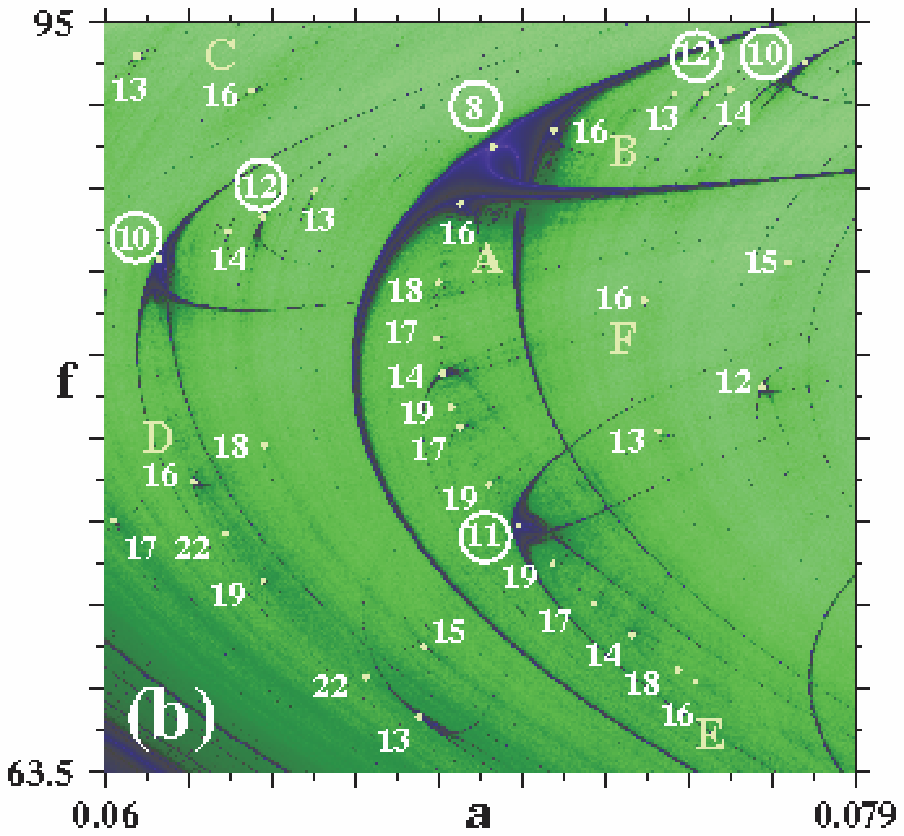}%
\includegraphics[width=5.7cm]{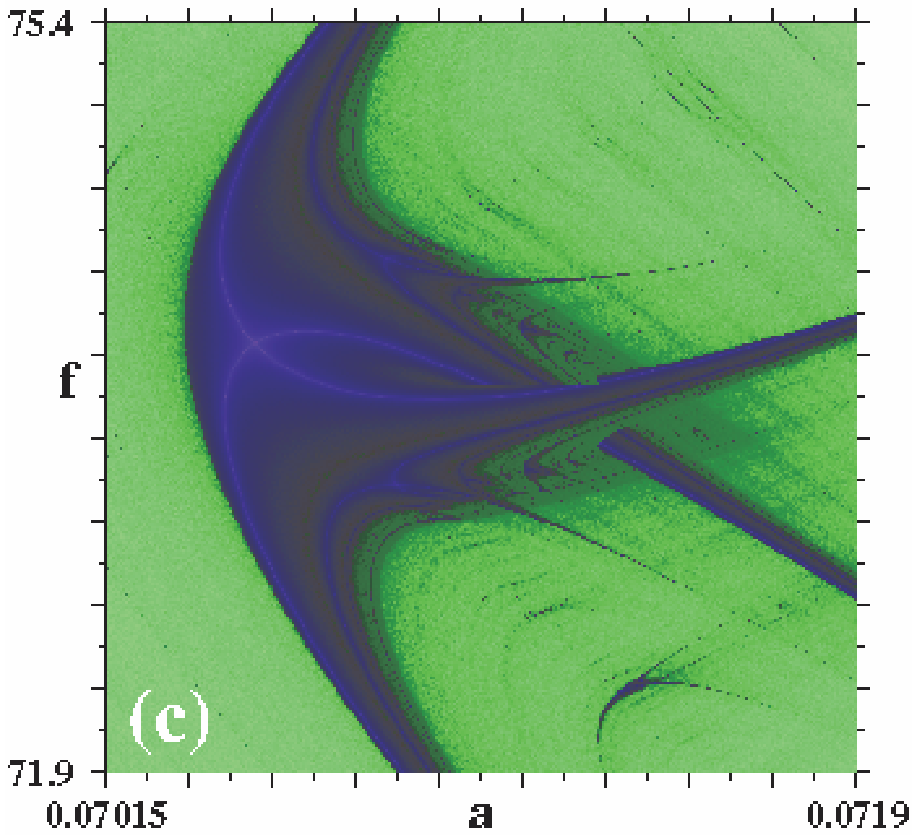}
\caption{\protect (Color online)
Structure of the frequency-amplitude phase-diagram of 
 the laser showing a regular network of stability islands 
(a)  Global view,
(b)  Zoom of the box in (a).
     Numbers indicate the main period of each stability island;
(c)  Magnification of the period-11 stability island (indicated in (b)
     by the encircled 11), displaying the generic shape of 
     all stability islands \cite{fest}.
 Color intensities are proportional to Lyapunov  exponents:
 blue for negative exponents (periodic oscillations), black for zero 
and green for positive exponents (chaotic oscillations).
  Frequencies are in kHz.}
\label{fig:fig1}
\end{figure*}


\begin{figure*}
\includegraphics[width=6.0cm]{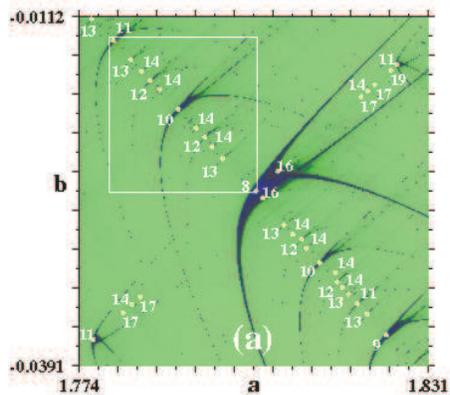}%
\caption{\protect (Color online)
Structure of the parameter space of the H\'enon map.
(a) The organization of shrimps here coincides with that of the laser
(compare with Fig.~\ref{fig:fig1}b).
  The fine structure observed around period-8 here
  reproduces the laser period-8 sequence along the curve passing by the
  encircled numbers in Fig.~\ref{fig:fig1}b;
(b) Magnification of the box in (a);
(c) Magnification of the box in (b). 
      Numbers indicate the main period of each stability island.
    Points mark windows, not doubly superstable crossings.
\xxr{Download ALL figures high-res PDF: 
     http://www.if.ufrgs.br/$\sim$jgallas/jg\_papers.html}}
\label{fig:fig2}  
\end{figure*}


Lasers with modulated parameters are arguably among the
simplest and most accessible laser systems of interest for
applications in science and engineering and 
for theoretical investigations. The
intrinsic interest in practical applications and in
the nonlinear dynamics of modulated lasers has spurred a wide
range of studies after the remarkably 
influential work of Arecchi {\it et al.}~\cite{arecchi82} reporting 
the first measurement of subharmonic bifurcations, 
multistability, and chaotic behavior in a
Q-swit\-ched  CO$_2$ laser. 
Since then, CO$_2$ lasers
have been fruitfully exploited in many situations. 
Recent applications include studies of
stochastic bifurcations in modulated CO$_2$ laser \cite{lora2004},
multistability induced by periodic modulations \cite{chiz_pre64}.
Rich nonlinear response of CO$_2$ lasers with current modulation
and cavity detuning \cite{pisa-josaB}, and
self-focusing effects in nematic liquid crystals \cite{brugioni}.

In the last 20 years the CO$_2$ laser was
extensively studied theoretically, numerically and experimentally 
\cite{gilmore,solari,tredicce,dangoisse}, but 
focusing mainly on the characterization of
dynamical behaviors in phase-space for specific parameters.
While a detailed description of phase-space dynamics
is already available in the literature \cite{gilmore,asy,hilborn,strogatz}, 
no equivalent description exists for the parameter space, 
except for works by
Goswami \cite{gos_pla190} who investigated analytically
the first few period-doubling bifurcations 
for the Toda model of the CO$_2$ laser \cite{toda}.

The present Letter reports an investigation of the parameter space 
of a paradigmatic model of class B lasers, the CO$_2$ laser.
More specifically, we study a popular two-level model of a CO$_2$ laser 
with modulated losses, focusing on the global stability
of the laser with respect to the modulation, not the intensity.
The remarkable discovery reported here is that
stability islands of the {\it continuous}-time laser model
emerge organized in a very regular network of self-similar structures
called shrimps \cite{jason}, 
illustrated in Figs.~\ref{fig:fig1} and \ref{fig:fig2},
and previously known to exist only in the parameter space of
{\it discrete}-time dynamical systems \cite{jason,fest,nusse,brian}.
Thus far, all attempts to uncover shrimps in flows, i.e.~in continuous-time
dynamical systems modeled with sets  of differential equations,
have failed to produce them \cite{potsdam}.

The single-mode dynamics of the loss-modulated CO$_2$ laser involves two
coupled degrees of freedom and a time-dependent parameter which we write, 
as usual \cite{chiz_pre64,gilmore,tredicce},
\begin{subequations} \label{modelo}
\begin{eqnarray}
\centering
\frac{du}{dt} &=& \frac{1}{\tau}\,(z-k)u,\label{momo}\\
\frac{dz}{dt} &=& (z_0 - z)\gamma - u z.  \label{model}
\end{eqnarray}
\end{subequations}
Here, $u$ is proportional to the radiation 
density, $z$ and $z_0$ are the gain and unsaturated gain in
the medium, respectively, 
$\tau$ denotes the transit time of the light in the laser cavity,
$\gamma$ is the gain decay rate, and $k\equiv k(t)$ represents the
total cavity losses.
The losses are modulated periodically as follows,
\begin{equation}
        k(t) = k_0(1 + a\cos 2\pi f t),  \label{loss}
\end{equation}
where $k_0$ is the constant part of the losses
and $a$ and $f$, the amplitude and frequency of the modulation, are
the main bifurcation parameters.
The remaining parameters are fixed at  $\tau=3.5\times 10^{-9}$ s, 
$\gamma=1.978\times 10^5$ s$^{-1}$, $z_0=0.175$  and $k_0=0.1731$.
These are realistic values, used in recent theoretical and experimental
investigations \cite{chiz_pre64}.
Integrations were done using the standard fourth-order 
Runge-Kutta scheme with fixed time-step,  equal to $h=4\times 10^{-8}$.
Phase diagrams in $a\times f$ space are obtained by computing Lyapunov 
exponents for a mesh of $600\times600$ equally spaced parameters.
Starting from an arbitrary initial condition, we ``followed the 
attractor'' that is, after increasing parameters we initiated iterations 
using the last obtained values as the new initial conditions.
The largest exponents were codified into a bitmap with a continuous 
color scale ranging from the maximum positive (green) to  maximum 
negative (blue) exponents. Zero exponents were codified in black.
One of the exponents is always zero since it is simply related to the 
time evolution. 
Three illustrative bitmaps for the laser model are 
shown in Fig.~\ref{fig:fig1}.

Figure \ref{fig:fig1}a displays a global view of the parameter space. 
The most prominent features, the broad curved structures in 
Fig.~\ref{fig:fig1}a,  show that the parameter space of the 
laser model above, Eqs.~(\ref{modelo}a-b), agrees qualitatively quite well
with the description 
of Goswami \cite{gos_pla190} for the Toda model of the CO$_2$ laser.
For the parameters chosen, the
relaxation frequency of our laser model is $50$ kHz.
From  Fig.~\ref{fig:fig1}a it is possible to see that there is a minimum
amplitude threshold $a$ beyond which subharmonic bifurcations start to 
occur, corresponding to about $100$ kHz, the harmonic of the relaxation
frequency. In addition, for certain parameter values new stability domains
are created by saddle-node bifurcations, each of them undergoing then its
own cascade of period doublings. 
So, in certain parameter ranges more than one stable mode coexist, 
giving rise to generalized multistability. This feature may be recognized in 
Fig.~\ref{fig:fig1} from the apparent sudden discontinuities in the coloring,
due only to the impossibility of plotting two distinct colors in the same place.

\begin{figure}[th]
\includegraphics[width=8.5cm]{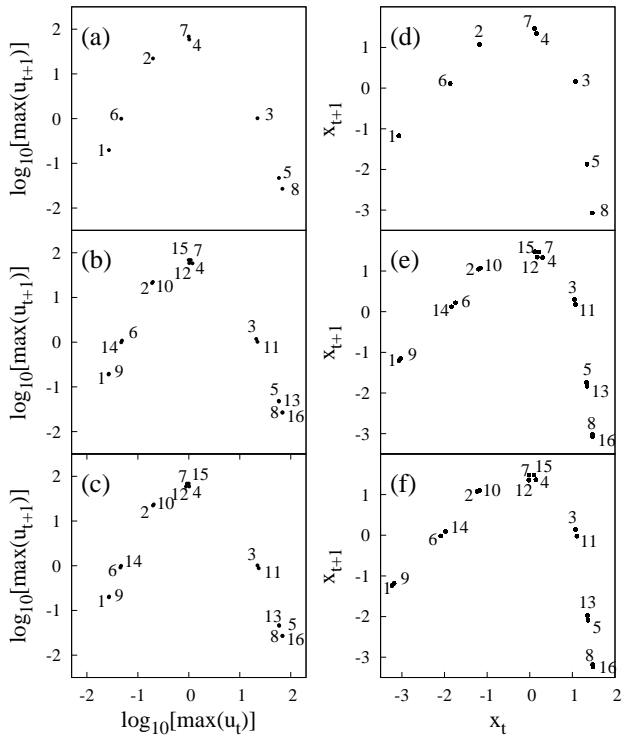}
\caption{\protect Comparison of return maps.
Left column: 
Laser return maps  
   for period-8 and its pair of doublings.
Parameters  are:
(a) $(a,f)=(0.06984$, $89.8)$, period 8,
(b) $(0.07138$, $90.47)$, period 16,
(c) $(0.06902$, $87.43)$, period 16.
   Frequencies are in kHz.
Right column: 
H\'enon return maps for period-8 and its doublings
seen at the center of Fig.~\ref{fig:fig2}a.
Parameters are:
(d) $(a,b)=(1.8087$,  $-0.02514)$, period 8,
(e) $(1.80395$, $-0.0257)$, period 16,
(f) $(1.80642$, $-0.02356)$, period 16.}
\label{fig:fig3}  
\end{figure}

The most interesting feature in Fig.~\ref{fig:fig1}a
is the remarkably regular structuring which appears 
in the region containing the box, shown magnified 
in Fig.~\ref{fig:fig1}b.
This figure  shows that embedded in the wide domain 
of parameters leading to chaotic laser oscillations there is a 
regular structuring of self-similar parameter windows, shrimps,
containing cascades of stable periodic oscillations, 
the {\it main period\/} of a few of the larger shrimps indicated 
by the number  near to them.
The period-11 shrimp seen in Fig.~\ref{fig:fig1}b is shown magnified in
Fig.~\ref{fig:fig1}c. Starting from the main period-11 body, it displays
two distinct doubling cascades as well as an infinite number of additional
period-doubling cascades, as thoroughly described for discrete-time systems
in Refs.~\cite{jason,fest}.

\begin{figure}[t]
\includegraphics[width=8.5cm]{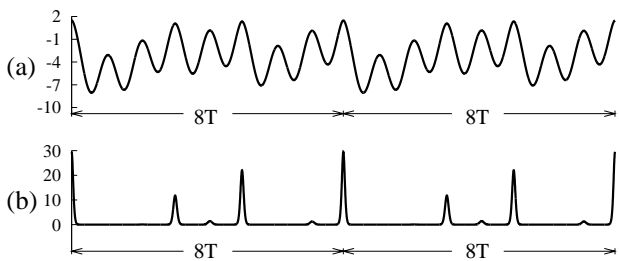}
\caption{Time evolution of the laser intensity $u(t)$ for the large
  period-8 structure in Fig.~\ref{fig:fig1}b, plotted in
 (a) logarithmic scale, 
 (b) linear scale. 
     Here $T=1/(89.8\hbox{ kHz})$ is the period of the modulation.}
\label{fig:fig4}  
\end{figure}

\begin{figure}[th]
\includegraphics[width=8.5cm]{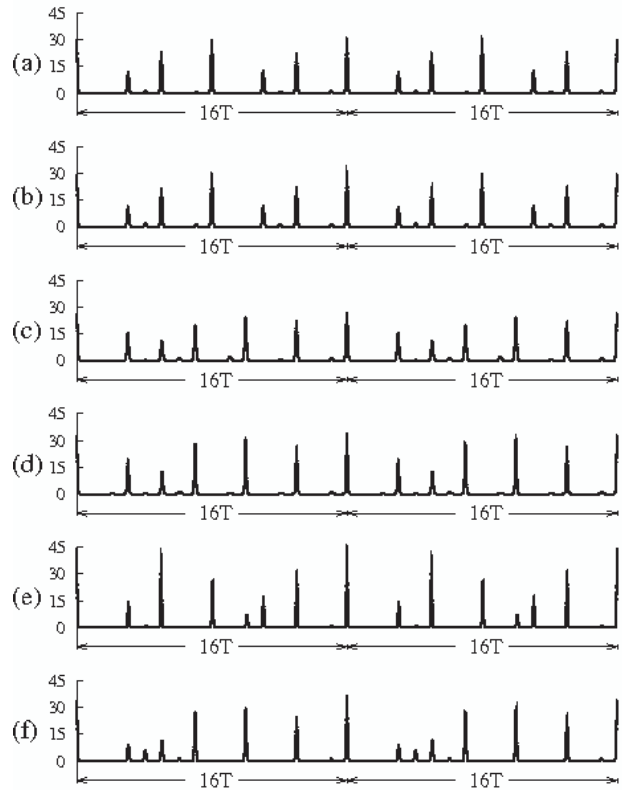}
\vspace{-2.0truecm}
\caption{\protect
Predicted signal intensity for the six period-16 stability islands 
labeled A, B, \dots F,
in Fig.~\ref{fig:fig1}b. Signals (a) and (b), at points A and B,
 are period-8 doublings.
All other signals are from islands which begin with period-16.
Note that signals look very similar, despite the fact that
they originate from very distinct regions of the parameter space.
Parameters $(a,f)$ are:
(a) $A=(0.06902$, $87.43)$,
(b) $B=(0.07138$, $90.47)$,
(c) $C=(0.063715$, $92.15)$, 
(d) $D=(0.062255$, $75.735)$,
(e) $E=(0.0749617$, $67.3281)$,
(f) $F=(0.073666$,  $83.359)$. Note that
 $T\equiv 1/f$ is slightly different for each signal.}
\label{fig:fig5}  
\end{figure}

The computation of bitmaps for the laser model is very computer demanding.
To alleviate this problem and to manifest the isomorphism between 
flows  and maps,
we display the generic fine and hyperfine structure of stability islands 
typically present in  multidimensional systems using the 
two-parameter H\'enon map as a paradigm:
\begin{equation}
 x_{t+1} = a-x_t^2 + b\,y_t, \qquad  y_{t+1} = x_t.  \label{henon}
\end{equation}
The nonlinearity
parameter $a$ (forcing) represents the bifurcation or control parameter. 
The damping parameter $b$ varies between $-1 \leq b \leq 1$, with 
$b=1$ representing the conservative limit and $b=0$ the limit of
strong damping.
While for $b=0$ there exists just a single chaotic attractor over a
wide range of the parameter $a$, several periodic and chaotic attractors 
coexist when $b\neq0$.

Since all these attractors evolve in the same generic manner, we consider 
here the strongly dissipative limit, focusing on slightly negative
values of $b$.
In this domain, Pando {\it et al.}~\cite{pando} found that a much 
more sophisticated
four-level model of the CO$_2$ laser with modulated losses 
behaves qualitatively similar to the H\'enon map.
The CO$_2$ laser dynamics, as of any class B laser, is 
characterized  by a time-delay between the intensity and the population 
inversion, a fact that nicely matches the delayed character of 
the H\'enon  map when written as a one-dimensional recurrence relation.

Figure \ref{fig:fig2} shows how regularly shrimps organize themselves
along very {\it specific directions\/} in parameter space.
The ordering along the main diagonal of Fig.~\ref{fig:fig2}a is 
the same found for the laser, in Fig.~\ref{fig:fig1}a, along 
the direction containing the encircled periods. 
Analogously, the secondary diagonal in 
Fig.~\ref{fig:fig2}a displays the same ordering that the 
parabolic arc in the middle of Fig.~\ref{fig:fig1}c.

The laser-H\'enon  agreement in parameter space permeates 
also to the phase space as corroborated by Fig.~\ref{fig:fig3}, 
comparing return maps for the laser (on the left column) 
with those for the H\'enon map (right column). 
The laser return maps were constructed using the sequence 
$u_\ell(t)$, $\ell=1, 2, 3, \dots$ of relative maxima of $u(t)$.
As it is easy to see, both sets of return maps agree remarkably
well \cite{elsewhere}.

How easy is to detect experimentally the regular structuring reported here?
Figure \ref{fig:fig4} illustrates a representative laser signal in two scales.
Although waveforms and underlying periodicities are easy to recognize 
in logarithmic scale, their experimental detection may become strenuous,
particularly as the period increases.
For instance, contemplating the six period-16 stability islands
in Fig.~\ref{fig:fig1}b, 
two of them arising from period-8 via saddle-node bifurcations,
one may ask what sort of differences should be expected in 
their measurement. The answer is depicted in Fig.~\ref{fig:fig5}.
In a real experimental setup, the difficulties to surmount are mainly to
access narrow high-period windows, and to have a wide enough
detection range. Modulated losses are usually obtained with an
intracavity polarizer and an electro-optical modulator. 
Recent progress in low-voltage electro-optical modulators
have considerably improved their performances \cite{eom}.
To detect large and small peaks simultaneously one can use 
a logarithmic preamplifier \cite{lefranc}.
Thus, detection and discrimination of the laser signals in 
Fig.~\ref{fig:fig5} is experimentally feasible with existing technology.

To uncover isomorphisms between  continuous-time (flows) and 
discrete-time (maps) in dynamical systems is an important event
from a physical and dynamical point of view and immediately raises 
interesting questions.
For instance, is an isomorphism to be expected also for
more refined laser models such that, e.g.,
of  Ciofini {\it et al.}~\cite{politi_model},
involving two rate-equations derived for a single-mode CO$_2$ laser
using center manifold theory for a four-levels model \cite{meucci}?
Their model is interesting because, as they say, it 
agrees well with experiments.
Do networks of shrimps exist for the laser model proposed 
very recently by Meucci {\it et al.}~\cite{meucci2004}. 
Do they exist for other laser models too?
Is it possible to find them in the parameter space of the 
laser parameters, not modulation? In autonomous systems?
Which laser system presents more restricted amplitude variations, 
making life easier for experimentalists?
What sort of mechanism generates shrimps in differential equations?
How are basins of attraction entangled in multistable domains
where self-similar structures are abundant \cite{rech}?
We will report on this elsewhere.

\medskip

We thank P.~Glorieux for a critical reading of the manuscript
and helpful suggestions.
CB thanks Conselho Nacional de Desenvolvimento Cient\'{\i}fico e
Tecnol\'ogico (CNPq), Brazil, for a doctoral fellowship.
JACG thanks CNPq for a senior research fellowship and the
Universit\'e de Lille for a ``Professeur invit\'e'' fellowship.



\begin{thebibliography}{99}

\bibitem{arecchi82} F.T.~Arecchi, R.~Meucci, G.~Puccioni, and 
  J.~Tredicce, 
      Phys.~Rev.~Lett. {\bf 49}, 1217 (1982).

\bibitem{lora2004} L.~Billings,  
 I.B.~Schwartz, D.S.~Morgan,
  E.M.~Bollt, R.~Meucci and E.~Allaria, 
  Phys. Rev. E {\bf70}, 26220 (2004).

\bibitem{chiz_pre64} V.N.~Chizhevsky, 
    Phys.~Rev.~E {\bf 64}, 036223 (2001).

\bibitem{pisa-josaB} A.N.~Pisarchik and B.F.~Kuntsevich,
J.~Opt.~B: Quantum Semiclass.~Opt.~{\bf 3}, 363 (2001).

\bibitem{brugioni} S.~Brugioni and R.~Meucci,
 Eur.~J.~Phys.~D, {\bf 28}, 277 (2004).


\bibitem{gilmore} R.~Gilmore and M.~Lefranc,
  {\it The Topology of Chaos, Alice in Stretch and Squeezeland},
  (Wiley, New York, 2002); R.~Gilmore,
   Rev.~Mod.~Phys. {\bf 70}, 1455 (1998).


\bibitem{solari} H.G.~Solari, E.~Eschenazi, R.~Gilmore, 
  and J.R.~Tredicce, 
Opt.~Commun.~{\bf 64}, 49 (1987).


\bibitem{tredicce} J.R.~Tredicce, F.T.~Arecchi, G.P.~Puccioni, 
  A.~Poggi, and W.~Gadomski, 
Phys.~Rev.~A {\bf 34}, 2073 (1986). 

\bibitem{dangoisse} D.~Dangoisse, P.~Glorieux, and D. Hen\-ne\-quin, 
Phys. Rev. A {\bf 36}, 4775 (1987);
Phys. Rev. Lett. {\bf 57}, 2657 (1986).

\bibitem{asy}  E.~Ott, {\it Chaos in Dynamical Systems}, 2nd edition,
  (Cambridge University Press, Cambrigde, 2002).
K.T.~Alligood, T.D.~Sauer and J.A.~Yorke,
 {\it Chaos: an Introduction to Dynamical Systems},
  (Springer, New York, 1997);

\bibitem{hilborn} R.C.~Hilborn,
 {\it Chaos and Nonlinear Dynamics: 
      An Introduction for Scientists and Engineers}, 2nd edition,
 (Oxford University Press, Oxford, 2000).

\bibitem{strogatz} S.H.~Strogatz,
 {\it Nonlinear Dynamics and Chaos: 
     With Applications to Physics, Biology, Chemistry and Engineering},
  (Perseus, Cambridge MA, 1994).

\bibitem{gos_pla190} B.K.~Goswami, 
Phys.~Lett.~A {\bf 190}, 279 (1994);
Phys. Lett. A {\bf 245}, 97 (1998).

\bibitem{toda} G.L.~Oppo and A.~Politi, 
Z.~Phys.~B {\bf 59},   111 (1985).

\bibitem{jason} J.A.C.~Gallas, 
Phys.~Rev.~Lett. 
{\bf 70}, 2714 (1993);
Physica A {\bf 202}, 196 (1994).

\bibitem{fest} J.A.C.~Gallas, 
        Appl. Phys. B, {\bf 60}, S-203 (1995), special supplement,
        Festschrift Herbert Walther.

\bibitem{nusse} J.A.C.~Gallas and H.E.~Nusse,
J. Economic Behavior and Organization {\bf 29}, 447 (1996).

\bibitem{brian} 
 B.R. Hunt, J.A.C. Gallas, C. Grebogi, J.A. Yorke and H. Ko\c cak,
 Physica D {\bf 129}, 35 (1999).


\bibitem{potsdam} M.~Thiel, M.C.~Romano, W.~von Bloh, and J. Kurths
 kindly informed us to have recently found {\it shrimps} 
 while computing recurrence plots.

\bibitem{pando}
  C.L.~Pando, G.A.~Luna Acosta, R.~Meucci and M. Ciofini,
  Phys. Lett. A {\bf 199}, 191 (1995).

\bibitem{elsewhere} 
A prototypical map particularly suited to investigate ana\-ly\-tically
the inner structure of stability islands is the 
{\it canonical\/} quartic map  $x_{t+1} = (x_t^2 -a)^2 -b$,
introduced in Ref.~\cite{jason} and
discussed at length in Refs.~\cite{fest,brian}.

\bibitem{eom} V.~Berger, N.~Vodjdani, D.~Delacourt and J.P.~Schnell,
 Appl. Phys. Lett. {\bf 68}, 1904 (1996).

\bibitem{lefranc} M.~Lefranc, D.~Hennequin, and P.~Glorieux,
Phys. Lett. A {\bf 163}, 269 (1992).

\bibitem{politi_model} M.~Ciofini, A.~Politi, and R.~Meucci,
Phys.~Rev.~A {\bf 48}, 605 (1993).

\bibitem{meucci} C.L.~Pando, R.~Meucci, M.~Ciofini, and F.T.~Arecchi,
      Chaos {\bf 3}, 279 (1993).

\bibitem{meucci2004} R.~Meucci, D.~Cinotti, E.~Allaria,
  L.~Billings, I.~Triandaf, D.~Morgan and I.B.~Schwartz,
  Physica D {\bf 189}, 70 (2004).

\bibitem{rech}  P.C.~Rech, M.W.~Beims  and  J.A.C.~Gallas,
   Phys. Rev. E {\bf 71},  017202 (2005).

\end{thebibliography}
\end{document}